\documentclass[fleqn,12pt,twoside]{article}
\usepackage[headings]{espcrc1}

\readRCS
$Id: espcrc1.tex,v 1.2 2004/02/24 11:22:11 spepping Exp $
\usepackage{graphicx}
\usepackage[figuresright]{rotating}

\newcommand{\AmS}{{\protect\the\textfont2
  A\kern-.1667em\lower.5ex\hbox{M}\kern-.125emS}}

\title{Thermalization of gluons at RHIC including $gg \leftrightarrow ggg$
interactions in a parton cascade}

\author{Z. Xu\address[MCSD]{Institut f\"ur Theoretische Physik,
        Johann Wolfgang Goethe Universit\"at Frankfurt, \\
        Max-von-Laue Str. 1, D-60438 Frankfurt, Germany}%
        \thanks{Email address: xu@th.physik.uni-frankfurt.de} and
        C. Greiner\addressmark\thanks{Email address:
        carsten.greiner@th.physik.uni-frankfurt.de}}
       
\runtitle{Thermalization of gluons at RHIC including $gg \leftrightarrow ggg$
interactions}
\runauthor{Z. Xu and C. Greiner}

\begin{document}

\maketitle

\begin{abstract}
Employing a newly developed pQCD inspired parton cascade we simulate 
the space time evolution of gluons which are produced initially in a heavy 
ion collision at RHIC energy. The inelastic $gg \leftrightarrow ggg$ 
interactions are for the first time implemented obeying full detailed 
balance. The numerical results show that thermalization of gluons is 
mainly driven by the inelastic gluonic interactions and reaches
equilibrium at $1\sim 2$ fm/c. In simulations for noncentral collisions
considerable partonic elliptic flow $v_2$ is generated being comparable 
with the experimental data.
\end{abstract}

\section{INTRODUCTION}

It was shown that the measured momentum anisotropy parameter $v_2$
at RHIC energy can be well described by ideal hydrodynamics \cite{kolb01}.
This indicates that the quark-gluon matter produced seems to behave like
a perfect fluid which represents a strongly interacting and thus locally
thermal system. On the other hand, the initial situation of the 
quark-gluon system is far from thermal equilibrium. It is therefore
important to understand how and which partonic interactions can thermalize
the system within a short timescale.

A convenient way to study thermalization of particles is to carry out
microscopical transport simulations which, however, need large
computational power. There are several such numerical realizations
\cite{zhang98} currently applied for investigating the space time
evolution of partons. In these models only elastic $gg \leftrightarrow gg$
interactions are considered and no thermal equilibrium can be realized
in Au+Au collision at RHIC when using reasonable pQCD cross sections.
It is thus essential to study the contribution of multiple interactions
to the thermal equilibration. In addition, the possible importance of
the inelastic scatterings on thermalization was raised in the ``bottom
up thermalization'' picture \cite{baier01}. Recently we have developed
a new $3+1$ dimensional Monte Carlo cascade solving the kinetic on-shell
Boltzmann equations for partons including inelastic
$gg \leftrightarrow ggg$ pQCD processes \cite{xu05}. Detailed balance
is fulfilled in a consistant manner.

The three-body gluonic interactions are described by the matrix element
\cite{biro93}
\begin{equation}
\label{m23}
| {\cal M}_{gg \to ggg} |^2 = \left ( \frac{9 g^4}{2} 
\frac{s^2}{({\bf q}_{\perp}^2+m_D^2)^2} \right ) 
\left ( \frac{12 g^2 {\bf q}_{\perp}^2}
{{\bf k}_{\perp}^2 [({\bf k}_{\perp}-{\bf q}_{\perp})^2+m_D^2]} \right )
\Theta(k_{\perp}\Lambda_g-\cosh y) \, ,
\end{equation} 
where $g^2=4\pi\alpha_s$. ${\bf q}_{\perp}$ and ${\bf k}_{\perp}$ denote,
respectively, the perpendicular component of the momentum transfer and 
that of the momentum of the radiated gluon in the c.m. frame of the 
collision. We regularize the infrared divergences by using the Debye
screening mass $m_D^2$ which is calculated locally over the actual
particle density obtained from the simulation. The suppression of
the radiation of soft gluons due to the Landau-Pomeranchuk-Migdal effect,
which is expressed via the step function in Eq. (\ref{m23}), is modeled
by the consideration that the time of the emission,
$\sim \frac{1}{k_{\perp}} \cosh y$, should be smaller than the time
interval between two scatterings or equivalently  the gluon mean free
path $\Lambda_g$. This leads to a lower cutoff of $k_{\perp}$ and
to an effective increase of the collision angles.

Until now only gluonic dynamics is considered in the cascade. In the
future quarks will be included. A special interest is put on the
investigation of elliptic flow of heavy quarks \cite{teaney05}.

\section{INITIAL CONDITIONS}

The production of the primary partons at the very onset of a heavy
ion collision is based on the picture of a free superposition of
minijets being liberated in the individual semihard nucleon-nucleon
interactions. Minijets denote here on-shell partons with transverse
momentum being greater than a certain cutoff $p_0$. Their production
is controlled by perturbative QCD for sufficient high $p_0$ \cite{wang91}.
On the other side, the smaller the cutoff $p_0$ is, the denser will be
the initial minijet system, which may accelerate thermalization.
Phenomenologically the cutoff $p_0$ can be chosen in a way to fit
the (final) $dE_T/dy$ as seen in experiment. The space time configuration
of the produced partons will be determined by applying the Glauber symmetry
with a Woods-Saxon nuclear distribution.

We have also considered the conditions of the initial partons according
to color glass condensate \cite{mv94}. First results can be found
in \cite{carsten05}.

\section{RESULTS}

\begin{figure}
\includegraphics[scale=0.74]{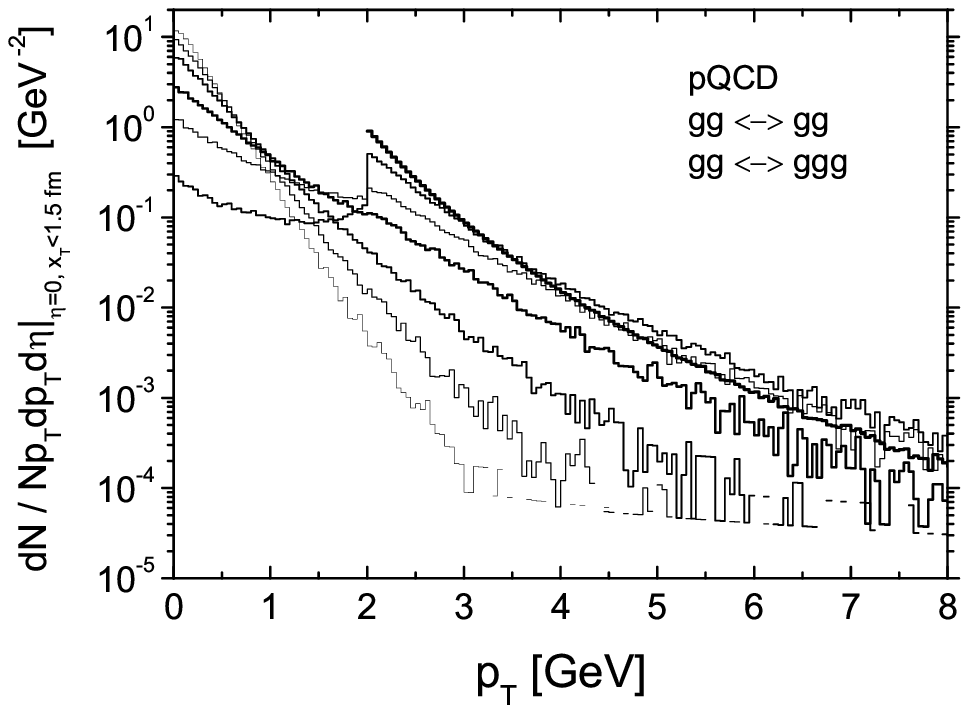}
\hspace{\fill}
\includegraphics[scale=0.74]{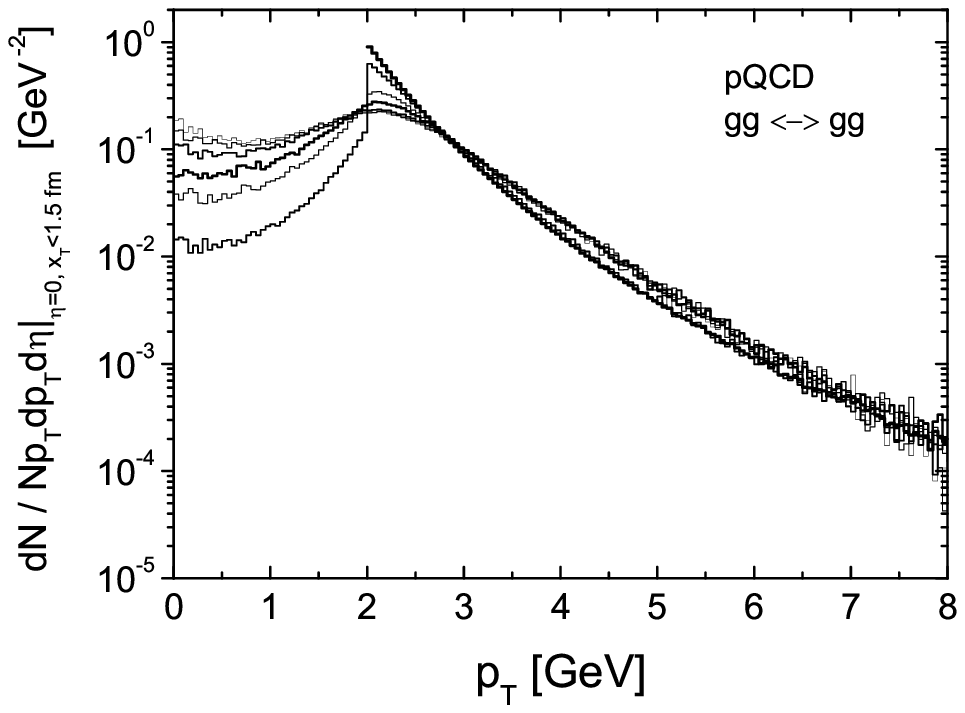}
\caption{Transverse momentum spectrum in the central region at
different times obtained from the simulation
with $gg \leftrightarrow ggg$ (left panel) and from the simulation 
without $gg \leftrightarrow ggg$ collisions (right panel).}
\label{dndpt}
\end{figure}

To study possible thermalization of gluons we concentrate on the local
central region which is taken as an expanding cylinder with a radius
of $1.5$ fm and within a unit interval of space time rapidity $\eta$
around the collision center $\eta=0$. Figure \ref{dndpt} shows the
varying transverse momentum spectrum with time obtained in the central
region.
The bold-faced histogram with a lower cutoff at $p_0=2$ GeV (a very
conservative setting) denotes the spectrum of the primary gluons
(or minijets). In the simulation including inelastic
$gg \leftrightarrow ggg$ scatterings (left panel of Fig. \ref{dndpt})
the curves from second upper to lowest depict, respectively, the spectrum
at $t=0.2$, $0.5$, $1$, $2$, $3$, and $4$ fm/c. We see that the
spectrum reaches an exponential shape at 2 fm/c and becomes increasingly
steeper at late times. This is a clear indication for the achievement
of thermal equilibrium and the onset of hydrodynamical evolution. In 
contrast, without including inelastic collisions (right panel of
Fig. \ref{dndpt}) one has no hints for equilibration.

\begin{figure}
\begin{minipage}[t]{75mm}
\includegraphics[scale=0.75]{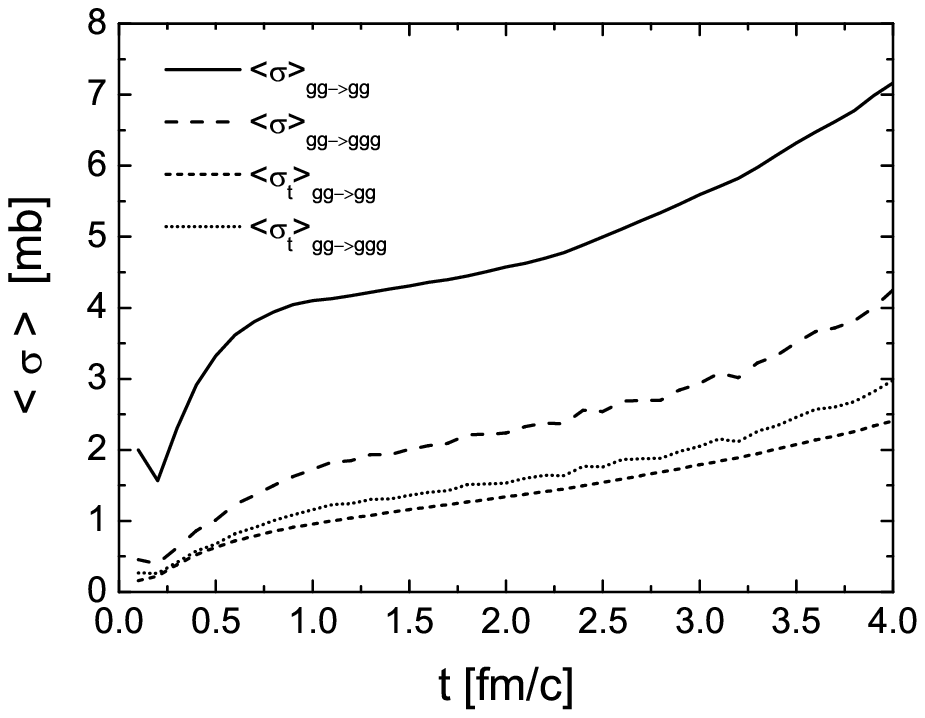}
\caption{Cross sections averaged in the central region. Results are 
obtained from the simulation including $gg \leftrightarrow ggg$ collisions.}
\label{cs}
\end{minipage}
\hspace{1mm}
\begin{minipage}[t]{75mm}
\includegraphics[scale=0.75]{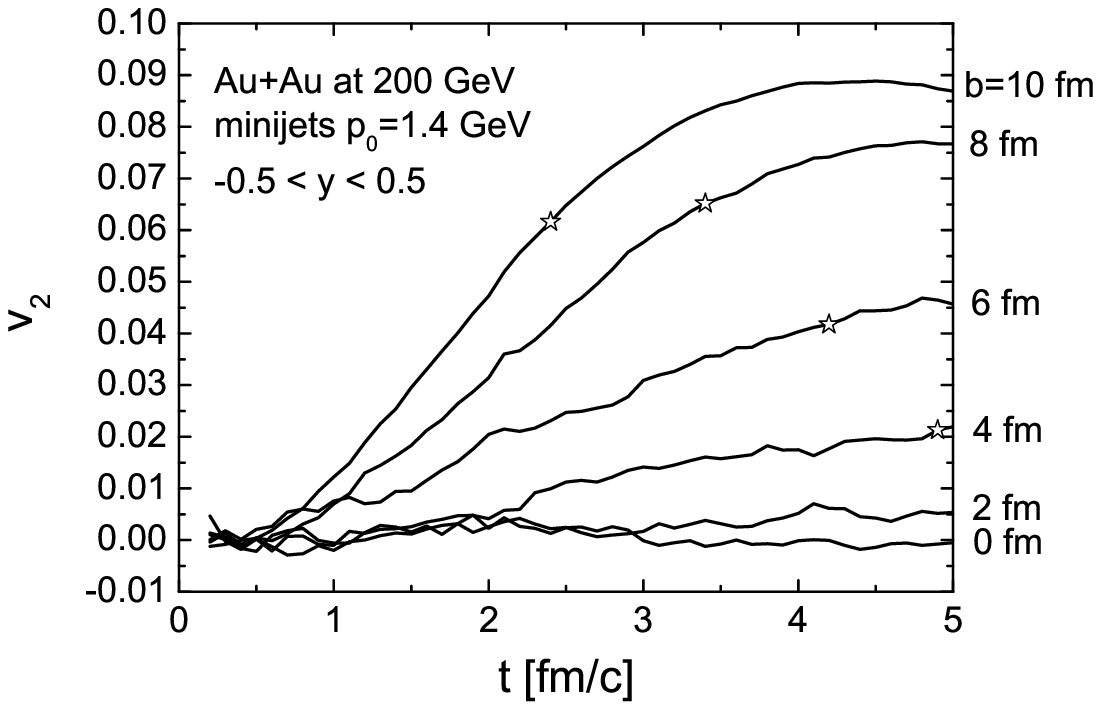}
\caption{Time evolution of the elliptic flow $v_2$ for various
impact parameter $b$.}
\label{v2}
\end{minipage}
\end{figure}

Figure \ref{cs} shows the cross sections of various gluonic
interactions and the corresponding transport cross sections. The latter
might be taken as a characteristic for momentum degradation.
While $\sigma_{gg \to gg}$ is significantly larger than
$\sigma_{gg \to ggg}$, its transport cross section is smaller than
that of typical $gg \to ggg$ collision. Inspecting the fraction of the
transport cross sections to the total cross sections we realize that the
distribution of the collision angle in $gg \to ggg$ processes is almost
isotropic, while the $gg \to gg$ collisions are much more forward peaked.
Taking the contribution of $ggg \to gg$ collision to the equilibration
into account, the total inelastic interactions are the dominant processes
compared to the elastic collisions. Besides the kinetic scatterings
plasma instabilities of the gluon field may also have contribution to
a very early and fast momentum equilibration \cite{instab}.
The latter should be further quantified.

We have also performed simulations when the momentum cutoff for the
initial minijets is taken smaller. It turns out that for the more dense
system at $p_0=1.3-1.5$ GeV (being in line with the measured $dE_T/dy$)
the full equilibrium comes slightly sooner at $1-2$ fm/c.
The timescale tends to saturate at smaller $p_0$. These results will be
presented in a sequent paper.

Taking $p_0=1.4$ GeV for the initial minijets we simulate the parton
evolution for noncentral collisions at RHIC energy in order to calculate
the elliptic flow parameter $v_2$. Figure \ref{v2} shows the time
evolution of $v_2$ extracted in the central rapidity for various impact
parameter $b$. These calculations are still preliminary and no exhaustive
tests have been finished. We see that $v_2$ increases with time and
saturates at late times, $3\sim 5$ fm/c. The larger the initial space
anisotropy is, the larger is the generated $v_2$. The results give us
strong indication that an early pressure is being built up. The symbols
in Fig. \ref{v2} mark the time from which the energy density in the
central region decreases below 1 $\mbox{Gev/fm}^3$. Therefore after
this time the system can be hardly described by the dynamics among partons.
If we take the $v_2$ values at the marked times as the contribution from 
the partonic phase, one realizes that they lie well in the region covered
by the experimental data with the systematic errors \cite{v2data}.

\end{document}